\documentclass[11pt]{article}
\usepackage{amssymb}
\usepackage{multicol}
\usepackage{amsmath}
\usepackage{longtable}

\newtheorem{example}{Example}[section]

\begin{document}
\today 
\begin{center} 
\begin{Large}
{\LARGE\bf 
The distribution of the maximum of a first order
moving average: the continuous case 
\\ http://arxiv.org/abs/0802.0523 
}\\[1ex]
\end{Large}
by\\[1ex]
Christopher S. Withers\footnote{Work begun while visiting the Statistics Dept, UNC, Chapel Hill.}\\
Applied Mathematics Group\\
Industrial Research Limited\\
Lower Hutt, NEW ZEALAND
\\[2ex] Saralees Nadarajah
\\ School of Mathematics\\ University of Manchester\\ Manchester M60 1QD, UK
\end{center}
\vspace{1.5cm}
{\bf Abstract:}~~We give the distribution of $M_n$,  the maximum of a
sequence of $n$ observations from a moving average of order 1.
Solutions are first given in terms of repeated integrals
and then for the case where the underlying independent random variables
have an absolutely continuous density.
When the
 correlation is positive ,
$$
P(M_n 
 \leq x)\ =\ \sum_{j=1}^\infty \beta_{jx} \ \nu_{jx}^{n}
\ \approx \ B_{x}\ \nu_{1x}^{n}
$$
where 
$\{\nu_{jx}\}$ are the eigenvalues (singular values) of a Fredholm kernel
and $\nu_{1x}$ is the eigenvalue of maximum magnitude.  
A similar result is given  when the correlation is negative.
The result is analogous to large deviations expansions for estimates, since
the maximum need not be standardized to have a limit.

For the continuous case the integral equations for the left and right eigenfunctions are 
converted to first order linear differential equations. 
The eigenvalues satisfy an equation of the form
$$\sum_{i=1}^\infty w_i(\lambda-\theta_i)^{-1}=\lambda-\theta_0$$
for certain known weights $\{ w_i\}$ and singular values  $\{ \theta_i\}$
of a given matrix.
This can be solved by truncating the sum to an increasing number of terms. 

\section{Introduction and Summary}
\setcounter{equation}{0}
Little is available in the literature on the behaviour of extremes of 
correlated sequences apart from some special cases involving Gaussian
processes and some weak convergence results. For example
Leadbetter et al. (1983) p59 give a convergence in distribution for
the scaled
maximum of a stationary sequence, and Resnick (1987) p239 gives
a similar result for moving averages.

This paper gives a powerful new method for giving the {\it exact}
distribution of extremes of $n$ correlated observations as weighted
sums of $n$th powers of associated eigenvalues. The method is
illustrated here for a moving average of order 1.

Let  $\{e_i\}$ be independent and identically distributed random variables
 from some distribution $F$ on $R$.
Consider the moving average of order 1,
\begin{eqnarray*}
X_i= e_i+ \rho e_{i-1} 
\end{eqnarray*}
where $\rho\neq 0.$
In Section 2 we give expressions for the distribution of the maximum
$$M_n=\max_{i=1}^n X_i$$
in terms of repeated integrals.
This is obtained via the recurrence relationship
\begin{eqnarray}
G_{n}(y) &=& I(\rho<0)G_{n-1}(\infty)F(y) +  {\cal K}  G_{n-1}(y)
 \label{joint} \\
\mbox{where }G_{n}(y) &=& P(M_n\leq x,\ e_n\leq y), \label{defG}
\end{eqnarray}
 $I(A)=1$ or 0 for $A$ true or false,
and
 $ {\cal K}$ is an  integral operator depending on $x$.
(Dependence on $x$ is suppressed.)
For this to work at $n=1$ we define
$$M_0=-\infty\mbox{ so  that  }G_{0}(y)=F(y).$$
In Section 3 we consider the case when $F$ is absolutely continuous 
 with density $f(x)$ with respect to Lebesque measure. 
In this case we show that corresponding to ${\cal K}$ is a Fredholm kernel
$K(y,z)$. We give a solution in terms of its eigenvalues and eigenfunctions. 
This leads easily to the asymptotic results stated in
the abstract.

Our expansions for $P(M_n\leq x)$ for fixed $x$ are large deviation results.
If $x$ is replaced by $x_n$ such that  $P(M_n\leq x_n)$ tends to the GEV
distribution, then the expansion still holds, but not the asymptotic expansion
in terms of a single eigenvalue, since this may approach 1 as $n\rightarrow \infty.$

Set $\int r=\int r(y)dy$.

\section{Solutions using repeated integrals.}
\setcounter{equation}{0}
  For $n\geq 1$, $G_n$ of (\ref{defG}) satisfies
\begin{eqnarray*}
G_{n}(y) &=& P(M_{n-1}\leq x, e_n+ \rho e_{n-1}\leq x, e_n\leq y)\\
&=&P(M_{n-1}\leq x,  e_{n-1}\leq (x-e_n)/\rho, e_n\leq y) \mbox{ if }\rho>0\\
 &=& \int^y  G_{n-1}((x-w)/\rho)dF(w)\\
&=&P(M_{n-1}\leq x,  e_{n-1}\geq (x-e_n)/\rho, e_n\leq y) \mbox{ if }\rho<0\\
 &=& \int^y  [G_{n-1}(\infty)- G_{n-1}((x-w)/\rho)]dF(w)
=G_{n-1}(\infty)F(y)- \int^y  G_{n-1}((x-w)/\rho)dF(w).
\end{eqnarray*}
(Thanks to the referee for noting a slip in the last line.)
That is, for $n\geq 1$, (\ref{joint}) holds with
\begin{eqnarray}
{\cal K} r(y) = \mbox{sign}(\rho) \int^y r((x-w)/\rho) dF(w).\label{calK}
\end{eqnarray}
Our goal is to determine
$$ u_n=P(M_n\leq x)= G_{n}(\infty).$$
In this section we give $u_n$ in terms of
\begin{eqnarray}
 v_n= [{\cal K}^n  F(y)]_{y=\infty}. \label{vn0} 
\end{eqnarray}
For example 
\begin{eqnarray}
v_1=-\int F(z)dF(x-\rho z) = -I(\rho<0)+\int F(x-\rho z) d F(z).\label{v1}
\end{eqnarray}
The behaviour of $u_n$ falls into two cases.

{\it The case $\rho>0$.} 
For $n\geq 1,$
\begin{eqnarray}
u_n=v_n \label{pos}
\end{eqnarray}
since
\begin{eqnarray*}
 G_{n}(y)= {\cal K}^n  F(y).
\end{eqnarray*}
 The marginal distribution of $X_1$ is $u_1=v_1$ given by (\ref{v1}).

{\it The case $\rho<0$.} 
By (\ref{joint}), for $n\geq 0,$
\begin{eqnarray}
G_{n+1}(y) &=& u_{n}F(y)+  {\cal K}G_{n}(y)\nonumber \\
 &=& a_n(y) \otimes u_n +a_{n+1}(y)
\mbox{ where }
a_i(y)={\cal K}^{i}F(y), \ a_n\otimes b_n=\sum_{j=0}^n a_jb_{n-j}. 
\label{2.4a} 
\end{eqnarray}
Putting $y=\infty$ gives the recurrence equation for $u_n$:
\begin{eqnarray}
 u_0=1,\ u_{n+1}=v_{n+1} +\sum_{i=0}^{n} v_iu_{n-i},\ n\geq 0. \label{rec} 
\end{eqnarray}
 The marginal distribution of $X_1$ is $u_1=1+v_1$ of (\ref{v1}).

{\bf An explicit solution for $u_n$ when $\rho<0$.}\\
Define the generating functions
\begin{eqnarray*}
U(t) =\sum_{n=0}^\infty u_nt^n,\ V(t)=\sum_{n=0}^\infty v_nt^n,
\end{eqnarray*}
Multiplying (\ref{rec}) by $t^n$ and summing from $n=0$ 
gives $(U(t)-V(t))/t=U(t)V(t)$, so that
\begin{eqnarray*}
U(t) &=&(1-W(t))^{-1}V(t)
\mbox{ where }W(t)=tV(t)=\sum_{n=1}^\infty w_nt^n,\ w_n=v_{n-1},\\
tU(t) &=&(1-W(t))^{-1}W(t)= (1-W(t))^{-1}-1=\sum_{j=1}^\infty  W(t)^j.
\end{eqnarray*}
By definition, for $j=0,1,\cdots$
$$W(t)^j=\sum_{n=j}^\infty \hat{B}_{nj}(w)t^n$$
where $\hat{B}_{nj}(w)$ is the {\it partial ordinary Bell polynomial} in
$w=(w_1,w_2,\cdots)$ tabled on p309 of Comtet (1974). For example 
$$
\hat{B}_{n0}(w)=\delta_{n0},\ \hat{B}_{n1}(w)=w_n,\ \hat{B}_{n1}(w)=w_1^n.
$$
So
$$tU(t)=(1-W(t))^{-1}-1=\sum_{n=1}^\infty \hat{B}_{n}(w) t^n$$
where 
$$\hat{B}_{n}(w)=\sum_{j=0}^n \hat{B}_{nj}(w)$$
is the {\it complete ordinary Bell polynomial}. For example 
$\hat{B}_{0}(w)=1.$
Taking the coefficient of $t^n$ gives the explicit solution
\begin{eqnarray}
u_{n-1}=\hat{B}_{n}(w),\ n\geq 1. \label{w2u} 
\end{eqnarray}
For example 
\begin{eqnarray*}
u_0 &=& \hat{B}_{1}(w)=\hat{B}_{11}(w)=w_1=v_0=1,\\
u_1  &=& \hat{B}_{2}(w)=\hat{B}_{21}(w)+\hat{B}_{22}(w)=w_2+w_1^2 =v_1+1.
\end{eqnarray*}
Similarly from  Comtet's table we can immediately read off $u_n, 1\leq n\leq 9$:
\begin{eqnarray*}
u_0 &=&1,\ u_1=v_1+1 \\
u_2 &=&v_2+2v_1+1, \\
u_3 &=& v_3+(2v_2+v_1^2)+3v_1+1,\\
u_4 &=&v_4+(2v_3+2v_1v_2)+(3v_2+3v_1^2)+4v_1+1, \\
u_5 &=& v_5+(2v_4+2v_1v_3+v_2^2) +(3v_3+6v_1v_2+v_1^3)+(4v_2+6v_1^2)+5v_1+1,\\
u_6 &=& v_6+(2v_5+2v_1v_4+2v_2v_3)+(3v_4+6v_1v_3+3v_2^2+3v_1^2v_2)+
(4v_3+12v_1v_2+4v_1^3)\\ && +(5v_2+10v_1^2)+6v_1+1,\\
u_7 &=& v_7+(2v_6+2v_1v_5+2v_2v_4+v_3^2)+(3v_5+6v_1v_4+6v_2v_3
+3v_1^2v_3+3v_1v_2^2 )\\ 
&& +
(4v_4+12v_1v_3+6v_2^2+12v_1^2v_2+v_1^4)+(5v_3+20v_1v_2+10v_1^3)+(6v_2+15v_1^2)+7v_1+1,\\
u_8 &=& v_8+(2v_7+2v_1v_6+2v_2v_5+2v_3v_4)+(3v_6+6v_1v_5+6v_2v_4+3v_3^2
+3v_1^2v_4+6v_1v_2v_3 +v_2^3)\\ &&+
(4v_5+12v_1v_4+12v_2v_3+12v_1^2v_3+12v_1v_2^2 +4v_1^3v_2)+(5v_4+20v_1v_3+10v_2^2+30v_1^2v_2+5v_1^4)\\ && +(6v_3+30v_1v_2+20v_1^3)+(7v_2+21v_1^2)+8v_1+1,\\
u_9 &=& v_9+(2v_8+2v_1v_7+2v_2v_6+2v_3v_5+v_4^2)+(3v_7+6v_1v_6+6v_2v_5
+6v_3v_4 +3v_1^2v_5+6v_1v_2v_4\\ && +3v_1v_3^2+3v_2^2v_3)+
(4v_6+12v_1v_5+12v_2v_4+6v_3^2 +12v_1^2v_4+24v_1v_2v_3+4v_2^3+4v_1^3v_3\\ &&
+6v_1^2v_2^2) +(5v_5+20v_1v_4
+20v_2v_3+30v_1^2v_3+30v_1v_2^2 +20v_1^3v_2+v_1^5)+(6v_4+30v_1v_3 +15v_2^2\\ &&
+60v_1^2v_2+15v_1^4)+(7v_3+42v_1v_2+35v_1^3)+(8v_2+28v_1^2)+9v_1+1.
\end{eqnarray*}
More generally {\it any} $u_n$ can be obtained from (\ref{w2u}) using the
recurrence relation 
\begin{eqnarray}
b_n=w_n\otimes b_n,\ n\geq 1, \mbox{ where }w_0=0,\
b_n=\hat{B}_{n}(w).
\label{bn} 
\end{eqnarray}
For example since $b_0=1$, this gives
$$b_1=w_1,\ b_2=w_1^2+w_2,\  b_3=w_1^3+2w_1w_2+w_3.$$
The recurrence relation (\ref{bn}) for the complete ordinary Bell polynomials
follows by taking the coefficient of $t^n$ in $(1-w)^{-1}-1=w(1-w)^{-1}$
where $w=W(t)$, and appears to be new.

\section{The absolutely continuous case.}
\setcounter{equation}{0}

Our solutions (\ref{pos}),  (\ref{rec}), (\ref{un}) do not tell us how
$u_n$ behaves for large $n$. Also they require repeated integration.
Here we give solutions that overcome these problems,
using Fredholm integral theory given in Appendix A.
Write (\ref{calK}) in the form
\begin{eqnarray}
{\cal K} r(y) = \int K(y,z) r(z)dz \mbox{ where }
 K(y,z) = \rho I(x\leq y+\rho z)f(x-\rho z). \label{kernel} 
\end{eqnarray}
Since
\begin{eqnarray}
  ||{\cal K}||_2^2=
 \int\int  K(y,z)K(z,y)dydz 
&=& \rho^2\int\int I(x<y+\rho z)I(x<z+\rho y)f(x-\rho z) f(x-\rho y)dydz 
\nonumber\\
&<&  \rho^2 \int\int f(x-\rho z) f(x-\rho y)dydz =1,\label{k2}
\end{eqnarray}
 $K(y,z)$ is said to be a Fredholm kernel w.r.t. Lebesgue measure, allowing
the Fredholm theory of the Appendix to be applied, in particular the 
functional forms of the Jordan form and singular value decomposition.
 If say, $0<\rho<1$, then one can show that
$$||{\cal K}||_2^2=\int F(x_t)dF(t)
\uparrow 1\mbox{ as }
x\uparrow \infty\mbox{  where }x_t=\min(x-\rho t,(x-t)/\rho).$$

Let  $\{\lambda_{j},r_{j},l_{j}:\ j\geq 1 \}$ be its eigenvalues (singular
values) and associated 
right and left eigenfunctions ordered so that 
$|\lambda_{j}|\leq |\lambda_{j+1}|.$
 By Appendix A these satisfy
\begin{eqnarray}
\lambda_{j} {\cal K} r_{j}=r_{j},\ 
\lambda_{j}  l_{j} {\cal K}=l_{j},\
\int r_{j}l_{k}=\delta_{jk},
  \label{lr} 
\end{eqnarray}
where $\delta_{jk}$ is the Kronecker function and we write
$\int a(y)b(y)dy=\int ab$. So $\{ r_{j}(y),l_{k}(y)\}$
are biorthogonal functions with respect to Lebesgue measure.
 Set
$$\nu_j= 1 /\lambda_j.$$
By (\ref{k2}) and (\ref{k22}),
 $$
1>||{\cal K}||_2^2=\sum_{j=1}^\infty \nu_j^{2}$$
where $\nu_j$ are the singular values, or if the Jordan form is diagonal,
the eigenvalues. (We shall use these terms interchangeably.)
So $|\nu_{j}|<1$ and $1+\nu_{j}>0$. 

Consider the case where the  Jordan form is diagonal.
Suppose that the eigenvalue $\lambda_1$ of smallest magnitude
has multiplicity $M$ (typically 1). Set
\begin{eqnarray}
 \beta_j= r_j(\infty)\int F l_j 
, \
B =\sum_{j=1}^M  \beta_j.
  \label{AB} 
\end{eqnarray}
 Then by (\ref{n}) for $n\geq 1$,
\begin{eqnarray}
v_n = \sum_{j=1}^\infty \beta_j \nu_j^n \label{vn} 
 = B \nu_1^{n}(1+\epsilon_{n})
 \label{deltav} 
\end{eqnarray}
where $\epsilon_{n}\rightarrow 0$ exponentially as $n\rightarrow \infty$.
(In fact by (\ref{n}) $1=v_0=\sum_{j=1}^\infty \beta_j$ if this converges.)
So for $n\geq 1,$ by (\ref{pos})
\begin{eqnarray}
 \mbox{ for }\rho>0,\ u_n= \sum_{j=1}^\infty \beta_j \nu_j^n.
 \label{pos2}
\end{eqnarray}
$\nu_1$ is given by (\ref{1}) with $\mu$ Lebesgue measure.
(When $\rho=0$ then (\ref{pos2}) holds with $\beta_j=\delta_{j1},\ \nu_1=F(x)$.
So we expect that $\nu_1\rightarrow F(x)$ as $\rho\downarrow 0.$)

Now suppose that  $\rho<0$. 

By (\ref{vn}), for $\max_{j=1}^\infty |\nu_j t|<1$,
$V(t)=1+\sum_{j=1}^\infty \beta_j\nu_jt /(1-\nu_jt)$.
So 
\begin{eqnarray*}
1-tV(t) &=& 1-t-\sum_{j=1}^\infty \beta_j \nu_jt^2/(1-\nu_jt)=N(t)/D(t)\\
\mbox{where }
D(t) &=& \Pi_{j=1}^\infty (1-\nu_jt),\
N(t)=\Pi_{j=1}^\infty (1-w_jt)\mbox{ say.}
\end{eqnarray*}
$D(t)$ is the Fredholm determinant of $K(x,y)$. (Now $w_j$ takes on
a different meaning than in Section 2.)
 So by the partial fraction expansion,
assuming that $\{w_j\}$ are all different,
\begin{eqnarray}
N(t)^{-1} &=& \sum_{j=1}^\infty c_j^{-1}(1-w_jt)^{-1}
\mbox{where }c_j = \Pi_{k\neq j}(1-w_k/w_j),\nonumber \\
 &=& \sum_{n=0}^\infty N_nt^n
\mbox{ where }
N_n=\sum_{j=1}^\infty c_j^{-1} w_j^n. \label{cN} 
\end{eqnarray}
Also by Fredholm's first theorem - see for example, p47 of Pogorzelski (1966),
$$D(t)=1+\sum_{n=1}^\infty D_n(-t)^n/n!,\
D_n=\int\cdots\int N{s_1\cdots s_n\choose s_1\cdots s_n}
ds_1\cdots ds_n 
$$
where $N{s_1\cdots s_n\choose s_1\cdots s_n}
=\det(N(s_j,s_k),1\leq j,k\leq n)$.
Alternatively, a simple expansion gives 
$$D_n/n!=\sum_{1\leq j_1<\cdots <j_n}\nu_{j_1}\cdots\nu_{j_n}=
[1^n],$$
the augmented symmetric function, in the notation of Table 10 of Stuart and 
Ord (1987). This table gives $[1^n]$ in terms of the power sums
$(r)=\sum_{j=1}^\infty \nu_j^r.$
For example $[1^3]=2(3)-(2)(1)+(1)^3$.
In our case
$$(r)=\int K_r(x,x)dx=\sum_{j=1}^\infty \nu_j^r$$
where  by (\ref{svd})
$$K_r(x,y)={\cal K}^{r-1}K(x,y)=\sum_{j=1}^\infty \nu_j^rr_j(x)l_j(y).$$
So  $[1^n]$ has the form
$$[1^n]=\sum_{k=1}^n \sum_{n_1+\cdots+n_k=n} A(n_1\cdots n_k)
(n_1)\cdots (n_k). $$ 
However this does not give its behaviour for large $n$.
At any rate, we have
$$1+tU(t)=(1-W(t))^{-1}=(1-tV(t))^{-1}=D(t)/N(t)$$
so that
\begin{eqnarray}
u_{n-1}= 
N_n\otimes D_n(-1)^n/n! \label{un} 
\end{eqnarray}
where $N_n$ is given by  (\ref{cN}). 
This solution will be useful for large $n$ if $D_n$ has an expansion
of the form
(\ref{vn}). However to date we have not been able to show this directly.
One can show that $D_n=(-1)^nB_n(d)$ where $B_n(d)$ is the complete exponential
Bell polynomial, $d_r=-(r-1)!w(n)$,  and $w(n)=\sum_{j=1}^\infty w_j^n=(n) $ for $w$. 
We conjecture that if $d_n=\sum_{j=1}^\infty a_j w_j^n$ where $|w_j|$
is strictly decreasing and $|a_j|>1$, then
$$B_n(d)\approx d_n\approx a_1w_1^n\mbox{ as }n\rightarrow\infty.$$

An alternative approach is to try a solution for $u_n$ of the form (\ref{vn}), say
\begin{eqnarray}
u_n = \sum_{j=1}^\infty \gamma_j \delta_j^n \label{un2} 
\end{eqnarray}
where $\delta_j$ decrease in magnitude.
Assuming that $\{\delta_j,\nu_j\}$ are all distinct, substitution into
the recurrence relation (\ref{rec}) gives us the following elegant
relations.  $\{\delta_j\}$ are 
the roots of
\begin{eqnarray}
\sum_{k=1}^\infty \beta_k/(\delta-\nu_k)=1\label{d} 
\end{eqnarray}
and $\beta_k$ is given by (\ref{AB}).
Having found  $\{\delta_j\}$ ,  $\{\gamma_j\}$ are the roots of
\begin{eqnarray}
\sum_{j=1}^\infty \gamma_j/(\delta_j-\nu_k)\equiv 1.\label{g} 
\end{eqnarray}
The last equation can be written 
$$A{\bf \gamma}={\bf 1}\mbox{ where }A=(A_{kj}: k,j\geq 1),\
 A_{kj}=1/(\delta_j-\nu_k)
.$$
So a formal solution is
$${\bf \gamma}=A^{-1}{\bf 1}.$$
Numerical solutions can be found by truncating the infinite matrix $A$ 
and infinite vectors ${\bf 1, \gamma}$
to $N\times N$ matrix and $N$-vectors, then increasing $N$ until the 
desired precision is reached.

{\bf Behaviour for large $n$ and $x$ independent of $n$.}\\
For $\rho>0$, (\ref{pos2}) implies
\begin{eqnarray}
 u_n
\approx  B \nu_1^{n}  \mbox{ and }\nu_1>0,
 \label{pos3}
\end{eqnarray}
where $B$ is given by (\ref{AB}).
Also $\nu_1$ is given by (\ref{1}) with $\mu$ Lebesgue measure.

Now suppose that $\rho<0$. By (\ref{un2}), 
\begin{eqnarray}
u_n =  \gamma_1 \delta_1^n(1+\epsilon_n') \label{unlarge} 
\end{eqnarray}
where $\epsilon'_{n}\rightarrow 0$ exponentially as $n\rightarrow \infty$
and $\delta_1$ has the largest magnitude among $\{\delta_n\}$.
(The case where multiple $\delta_n$ exist of magnitude  $|\delta_1|$
requires an obvious adaptation.)

{\bf Integral and differential equations for the eigenfunctions and
resolvent.}\\
The  right eigenfunctions satisfy 
$\nu_j r_{j}={\cal K}r_{j}$, that is
\begin{eqnarray}
\nu_j r_{j}(y)= {\cal K} r_{j}(y)
=\mbox{sign}(\rho) \int^y r_{j}((x-w)/\rho)dF(w) \label{ri}
\end{eqnarray}
For example 
$$\nu_j r_j(\infty)=\rho\int r_j(z)f(x-\rho z)dz.$$
Differentiating gives the non-standard linear first order  
differential equation
\begin{eqnarray}
 \nu_j\dot{r}_{j}(y)
=\mbox{sign}(\rho) f(y)r_{j}((x-y)/\rho),\ r_{j}(-\infty)=0.\label{de}
\end{eqnarray}
 Similarly
the left eigenfunctions satisfy \\
$\nu_jl_{j}=l_{j}{\cal K}$, that is
\begin{eqnarray}
\nu_jl_{j}(z)=\int l_j(y)K(y,z)dy =\rho f(x-\rho z)\int_{x-\rho z} l_{j}(y)dy.
 \label{le}
\end{eqnarray}
 So
\begin{eqnarray}
l_{j}(-\infty)=0\mbox{ if }\rho> 0,\
l_{j}(\infty)=0\mbox{ if }\rho <0, \label{linf}
\end{eqnarray}
and by differentiating,
\begin{eqnarray}
\nu_j (d/dz)[l_{j}(z)/f(x-\rho z)]=\rho^2 l_{j}(x-\rho z). \label{lj}
\end{eqnarray}
The resolvent satisfies
\begin{eqnarray}
[K(y,z,\lambda)-K(y,z)]/\lambda ={\cal K} K(y,z,\lambda)= K(y,z,\lambda){\cal K}. \label{res}
\end{eqnarray}
So
\begin{eqnarray*}
K(-\infty,z,\lambda) &=&0, \\
K(y,\infty,\lambda) &=& 0\mbox{ if } \rho<0,\\
K(y,-\infty,\lambda) &=& 0\mbox{ if } \rho>0
\end{eqnarray*}
and by differentiation the resolvent satisfies the first order partial 
 differential equations
\begin{eqnarray*}
(\partial/\partial y)\mbox{ LHS}(\ref{res}) &=& \mbox{sign}(\rho) f(y)K((x-y)/\rho,z,\lambda),\\
(\partial/\partial z)[\mbox{ LHS}(\ref{res})/f(x-\rho z)] &=& \rho^2 K(y,x-\rho z,\lambda).
\end{eqnarray*}

These may involve the Dirac function $\delta(x)$ since with $x_z=x-\rho z$,
\begin{eqnarray*}
(\partial/\partial y) K(y,z) &=& \rho f(x_z)\delta(y-x_z),\\
(\partial/\partial z) K(y,z) &=&
 \rho^2 f(x_z)\delta(y-x_z) -\rho^2 I(x_z<y) \dot{f}(x_z).
\end{eqnarray*}
For special cases, it is possible to solve (\ref{ri}) or
(\ref{le}) explicitly.
\begin{example} 
Suppose that $F(y)=ae^{ay}$ on $(-\infty, 0]$ where $a>0$, and that
$\nu_j<0,\ \rho<-1,\ y\leq x.$ Taking $r_j(0)=1$, a solution of  (\ref{ri})
is
$$r_j(y)=e^{b_jy}\mbox{ where }b_j|\nu_j|/a=e^{b_jx/\rho},\ b_j>0.$$ 
\end{example}

{\bf Formal expressions for the eigenfunctions.}\\
We now give a formal  solution of (\ref{de}) for  $r_j$ in terms of $r_j(0)$.
(The value 0 is arbitrary: a similar solution can be obtained in terms of 
$r_j(y_0)$ for any $y_0$.)
Set $$ r(y)=r_j(y),\ c=\lambda_j \ \mbox{sign}(\rho).$$
Suppose that $f$ and $r$ have Taylor series
expansions about 0. Denote the $i$th derivatives of $f(y)$ by $f_{.i}(y)$   and
set $f_i=f_{.i}(0),\ r_i=r_{.i}(0)$. Expanding
$$\dot{r}(y)=cf(y) r((x-y)/\rho)$$
about 0, for $i\geq 0$ the coefficient of $y^i/i!$ is
 $$r_{i+1}=c\sum_{a+b=i} {i\choose a} f_a r_{.b}(x/\rho)(-\rho)^{-b}
=c\sum_{k=0}^\infty q_{ik} r_k=cf_i r_0 +c\sum_{k=1}^\infty q_{ik} r_k
$$
where
\begin{eqnarray}
 q_{ik} =
\rho^{-k}\sum_{b=0}^{\min(i,k)} {i \choose b} f_{i-b} (-1)^b x^{k-b}/(k-b)!.
 \label{q}
\end{eqnarray}
For $l,k\geq 1$ set $Q_{lk}= q_{l-1,k}$. Set $Q=(Q_{lk}:l,k\geq 1) $.
Set 
\begin{eqnarray}
{\bf f}'=(f_0,f_1,\cdots),\ {\bf R}'=(r_1,r_2,\cdots),\
 {\bf Y}_y'= {\bf Y}'=(y/1!, y^2/2!, \cdots).
 \label{Yy}
\end{eqnarray}
So 
${\bf R}={\bf f} cr_0+cQ{\bf R},\ {\bf R}=(I-cQ)^{-1} Fc r_0$.
But $r(y)-r(0)={\bf Y}'{\bf R}$. So we obtain the $j$th right eigenfunction in terms of its
value at 0:
$$r(y)/r(0)=1+{\bf Y}'(c^{-1}I-Q)^{-1} {\bf f},$$
that is,
\begin{eqnarray}
r_j(y)/r_j(0)=1+{\bf Y}'(d_j I-Q)^{-1} {\bf f} \mbox{ where } d_j=\nu_j \ \mbox{sign}(\rho)
. \label{r}
\end{eqnarray}
For example for the extreme value distribution $F(x)=e^{-e^{-x}},\
{\bf f}=e^{-1}(1,0,-1,-1,-7/288,-31/4,\cdots)'.$ 

Since $r_j$ is unique only up to a constant multiplier, we may take $r_j(0)\equiv 1.$
The solution (\ref{r}) can now be implemented by successive approximations. For $N\geq 1$ set
\begin{eqnarray}
r_{Nj}(y)/r_j(0)=1+{\bf Y}_N'(d_jI_N-Q_N)^{-1} {\bf f}_N \label{rN}
\end{eqnarray}
where ${\bf Y}_N, {\bf f}_N$ are the 1st $N$ elements of ${\bf Y},{\bf f}$ and $Q_N$ is the upper left
$N\times N$ elements of $Q$. 
Then one expects that $r_{Nj}(y)\rightarrow r_j(y)$ as $N\rightarrow \infty,$
giving the $j$th left eigenfunction.

 A similar treatment of (\ref{lj}) gives an equation for the $j$th left
eigenfunction in terms of its value at at an arbitrary point, taken here as
$x$. Set
$$c=\rho^2\lambda_j,\ l=l_j,\ e(y)=f(y)^{-1}.$$
By Taylor expansions,
$$l(z)e(x-\rho z)=\sum_{i=0}^\infty (z^i/i!) \sum_{a+b=i} {i\choose a} l_{.a}(0)e_{.b}(x)(-\rho)^b.$$
By (\ref{lj}), $l$ satisfies $(d/dz) \mbox{LHS}=cl(x-\rho z).$
Taking the coefficient of $z^i/i!$, for $i\geq 0$,
\begin{eqnarray}
 \sum_{a+b=i+1} {i+1\choose a} l_{.a}(0)e_{.b}(x)(-\rho)^b
=cl_{.i}(x) (-\rho)^i. \label{oh}
\end{eqnarray}
By another Taylor expansion,
$$ l_{.a}(0)=\sum_{k=0}^\infty l_{.k+a}(x) (-x)^k/k!.$$
So LHS of (\ref{oh}) is 
$\sum_{j=0}^\infty W_{ij}  l_{.j}(x)= V_i l(x)+(W{\bf L})_i$
where we set
\begin{eqnarray}
  W_{ij} &=& \sum_{a=0}^{\min(j,i+1)} {i+1\choose a} e_{.i+1-a}(x)
(-\rho)^{i+1-a} (-x)^{j-a}/(j-a)!,\nonumber \\
 W &=&( W_{ij}:\ i,j\geq 1),\
U_i = W_{i0}=e_{.i+1}(x)(-\rho)^{i+1},\ V_j= W_{0j},\nonumber \\
 L_j &=&l_{.j}(x),\ {\bf L}'=(L_1,L_2,\cdots),
\ D_r=diag( r^i:\ i\geq 1), \ r=-\rho.
 \label{W}
\end{eqnarray}
So (\ref{oh}) for $i\geq 1 $ can be written 
$ {\bf U} l(x)+W{\bf L}=cD_r {\bf L} $
so that
 ${\bf L}=(cD_r -W)^{-1}{\bf U} l(x) $ giving in the notation of (\ref{Yy}),
$$ l(z)-l(x)={\bf Y}_{z-x}'{\bf L}={\bf Y}_{z-x}'(cD_r -W)^{-1}{\bf U} l(x). $$
 That is,
$ l_j(z)=l(z)$ is given by
\begin{eqnarray}
 l(z)/l(x) = 1+{\bf Y}_{z-x}'(cD_r -W)^{-1} {\bf U}.\label{l}
\end{eqnarray}
Finally, the value of the multiplier $l_j(x)=l(x)$ is determined by (\ref{bi}):
$$1/l_j(x)=\int r_j(z) RHS(\ref{l})dz.$$
{\bf An equation for the eigenvalues.}\\
Substituting into  (\ref{oh}) at  $i=0  $, that is   
$ W_{00}l(x)+{\bf V}'{\bf L}=cl(x),$
 we obtain
 \begin{eqnarray}
{\bf V}'(cD_r -W)^{-1}{\bf U}=c-W_{00}.
\label{ev}
\end{eqnarray}
The roots  $c$ of this equation are just $\{\rho^2\lambda_j\}$, so this is the
  equation for the eigenvalues we have been seeking.
If $\{\ \theta_j,\ j\geq 1\}$ are the singular values of $D_r^{-1}W$,
and if this has diagonal Jordan form $D_r^{-1}W=R_0\Lambda L_0^*$
where $\Lambda=\mbox{diag}(\theta_1,\theta_2,\cdots),$
(see (\ref{DJF}) below), 
then (\ref{ev}) can be written
$$\sum_{i=1}^\infty w_i(c-\theta_i)^{-1}=c-\theta_0,$$
where now the weights $\{ w_i\}$ are given by
$$ w_i=v_iu_i \mbox{ where }
{\bf v}=\bar{R}_0D_r^{-1}{\bf V},\ {\bf u}= L_0^*{\bf U}.$$
 If  $\{c_{Nj:\ j=1,\cdots,N+1}\}  $ are the roots of its $N  $ dimensional approximation, say
 \begin{eqnarray*}
 {\bf V}_N'(cD_{\rho N} -W_N)^{-1}{\bf U}_N=c-W_{00},
\label{evN}
\end{eqnarray*}
then  $c_{Nj}\rightarrow c_j=\rho^2\lambda_j  $ as $N\rightarrow \infty.  $ 
(This is essentially a polynomial in $c$ of degree $N+1$.)
Having obtained an eigenvalue, one can substitute it into (\ref{r})
 and (\ref{l}) to obtain
the corresponding eigenfunctions up to constants $l(x)$ and $r(0)$.
 As noted
in the appendix, either of these (but not both) can be arbitrarily chosen. 
The conditions $r(-\infty)=0$ and (\ref{linf}) can be verified numerically.
\begin{example}
Suppose that $f=\phi$, the density of a standard normal ${\cal N}(0,1)$ r.v..
Then $e_{.j}(x)=\phi(x)^{-1}H_j^*(x)$ where 
$$H_j^*(x)=E\ (x+{\cal N}(0,1))^j=\sum_k {j\choose 2k} x^{j-2k} m_{2k}$$
is the modified Hermite polynomial
and $ m_{2k}=(2k)!/k!2^k$ is the $2k$th  moment of ${\cal N}(0,1)$. See
  Withers and McGavin (2006).
\end{example}
 An alternative is to expand RHS(\ref{oh}) about $x=0$, giving
$c(-\rho)^i(l(0)+{\bf U}'{\bf L})$ where we set
\begin{eqnarray*}
L_j &=& l_{.j}(0),\ {\bf L}'=(L_1,L_2,\cdots),\
x_k=x^k/k!,\ {\bf U}'=(x_1,x_2,\cdots),\\
A_{ia} &=& l_{.a}(0)[e_{.b}(x)(-\rho)^b]_{b=i+1-a},\
A=(A_{ia}:\ i,a\geq 1),\ V_i=A_{i0}=[e_{.b}(x)(-\rho)^b]_{b=i+1}.
\end{eqnarray*}
Let $0^i$ denote the row $i-$vector of zeros.
For $i\geq 1$, (\ref{oh}) gives
$$V_il(0)+(A_{i1},\cdots,A_{i,i+1},0,0,\cdots){\bf L}
=c(-\rho)^i (0^{i-1},x_0,x_1,\cdots){\bf L}.$$
 That is
$${\bf V}l(0) +A{\bf L}=cD_r X{\bf L}$$
where the $i$th row of the matrix $X$ is $(0^{i-1},x_0,x_1,\cdots).$
So
$${\bf L}=B^{-1}{\bf V}l(0)\mbox{ where }B=cD_r X-A,\ 
l(y)/l(0)=1+ {\bf Y}_y'B^{-1}{\bf V}.$$
$X$ is upper triangular, while $A$ is lower triangular except for the 1st 
super-diagonal.
For $i=0$, (\ref{oh}) gives 
$$\sum_{a=0}^1 A_{0a} l_a=c\sum_{k=0}^\infty l_kx_k
=cl(0)+c{\bf U}'{\bf L}.$$
So we obtain as {\it an alternative equation for the eigenvalues}
$$ A_{00}  +A_{01} (B^{-1}{\bf V})_1=c+c{\bf U}' B^{-1}{\bf V}
\mbox{ where } A_{00}= -\rho e_{.1}(x),\ A_{01}=  e(x).
$$
Unfortunately Appendix A cannot be applied with $\mu=F$ since
\\ 
${\cal K}G(y)=\mbox{sign}(\rho)\int^y G((x-w)/\rho)dF(w)$
is not of the form $\int K(y,z)G(z)dF(z).$
It would be of great interest, and in particular allow a unified approach
to {\it this }problem, if Fredholm's theory can be extended to the system
$${\cal K} {\cal O} r=\nu r,\ {\cal K}^* {\cal O}^* l={\bar \nu} l,\
 l_i^*{\cal O} r_jd\mu=\delta_{ij}$$
for ${\cal K}$ an $q\times q$ integral operator with kernel
$K(y,z):\ R^p\times R^p\rightarrow C^{q\times q}$ with respect a measure $\mu$,
 ${\cal O}$ a  $q\times q$ operator,
where $*$ is the transpose of the complex conjugate, and  ${\bar \nu}$ is the
 complex conjugate of $ \nu$ .

For our problem, one could then apply the theory with 
$$p=q=1,\ \mu=F,\ K(y,z)=\mbox{sign}(\rho) I(z<y)
,\ {\cal O}G(w)=G((x-w)/\rho).$$

\appendix
\begin{center}
\section*{\normalsize APPENDIX A}
\end{center}
\addtocounter{section}{1}
\setcounter{equation}{0}

To make the paper self-contained, we give here some theory for 
Fredholm integral equations with non-symmetric kernels. 

First consider the case where $K$ is
any $k\times k$ complex matrix. Its singular value decomposition is
$$K=R\Lambda L^* \mbox{ where }RR^*=I,\ LL^*=I,\ 
\Lambda=\mbox{diag}(\nu_1,\nu_2,\cdots),
$$
* denotes the complex conjugate transpose.
Since
$$
KK^*R=R\Lambda\Lambda^*,\ K^*KL=L\Lambda^*\Lambda,
$$
the $j$th column of $R$ is a right eigenvector of $KK^*$ with eigenvalue 
$|\lambda_j|^2$ and
the $j$th column of $L$ is a right eigenvector of $K^*K$ with 
the same eigenvalue. If $K$ is non-singular, its inverse is
$$K^{-1}=L\Lambda^{-1} R^*.$$
If it is singular, a pseudoinverse is given by
$$K^{-}=L\Lambda^{-} R^*.$$
However the singular value decomposition does not give a nice form for powers 
of $K$.
This drawback is overcome by its Jordan decomposition.
Consider the case where this is diagonal.
Then
\begin{eqnarray}
K=R\Lambda L^* \mbox{ where }RL^*=I,\mbox{ and }
\Lambda=\mbox{diag}(\nu_1,\nu_2,\cdots,)
\label{DJF}
\end{eqnarray}
is composed of the eigenvalues of $K$.
Then for any complex $\alpha$,
$$K^\alpha=R\Lambda^\alpha L^*.$$
Taking  $\alpha=n$ and $\alpha=-1$ gives the $n$th power and inverse of $K$.

Now let $K(y,z)$ be a real function on $\Omega \times\Omega $ where
$\Omega$ is a subset of $R^p$. Suppose that $\mu$ is a $\sigma$-finite measure
on $\Omega$ and that
$$
0< ||{\cal K}||_2^2=
 \int\int  K(y,z)K(z,y)d\mu(y)d\mu(z)<\infty.
$$
(This $L_2$ condition can be changed to
$$
 \int  |K(y,y)|d\mu(y)<\infty
$$
at the expense of notational complexities that need not concern us here.)
The corresponding integral operator ${\cal K}$ is defined by
\begin{eqnarray}
{\cal K}\phi(y)=\int K(y,z) \phi(z) d\mu(z),\
\phi(z) {\cal K}
=\int  \phi(y) K(y,z) d\mu(y).
 \label{joint1} 
\end{eqnarray}
The Fredholm equations of the first kind,
$$
\lambda{\cal K}r(y)=r(y),\ 
\lambda l(z){\cal K}= l(z), 
$$
have only a countable number of solutions, say
  $\{\lambda_{j},r_{j}(y),l_{j}(z),\ j\geq 1 \}$
up to arbitrary constant multipliers for   $\{r_{j}(y),\ j\ge 1 \}$,
and these satisfy
\begin{eqnarray}
\int r_{j}l_{k}d\mu=\delta_{jk} \mbox{ where } 
\int gd\mu=\int_{R^p} g(y)d\mu(y).\label{bi} 
\end{eqnarray}

These are called the {\it singular values (or eigenvalues) and right and left eigenfunctions }
of $(K,\mu)$ or  ${\cal K}$.
Also
\begin{eqnarray}
 K(y,z)=\sum_{j=1}^\infty r_j(y)l_j(z)/\lambda_j \label{svd} 
\end{eqnarray}
with convergence  in $L_2(\mu\times\mu)$, or more strongly under other
conditions: see  Withers (1974, 1975, 1978).
This is the functional form of the Singular Value Decomposition for a square
non-symmetric matrix. 

 Fredholm equations of the second kind,
$$
r(y)-\lambda{\cal K}r(y)=f(y),\
 l(z)-\lambda l(z){\cal K}=g(z),
$$
can be solved for $\lambda$ not an eigenvalue using
$$ 
(I-\lambda{\cal K})^{-1}=I+\lambda{\cal K}_\lambda
$$
where
$${\cal K}_\lambda f(y)=\int K(y,z,\lambda)f(z)d\mu(z),\
g(z){\cal K}_\lambda=\int g(y)K(y,z,\lambda)d\mu(y),
$$
and the {\it resolvent} $K(y,z,\lambda)$ with operator ${\cal K}_\lambda$
is the unique solution of
$$ (I-\lambda{\cal K}){\cal K}_\lambda={\cal K}
={\cal K}_\lambda(I-\lambda{\cal K}),
$$
that is,
$$\lambda \int K(y,u) K(u,z,\lambda)d\mu(u)= K(y,z)- K(y,z,\lambda)
=\lambda \int K(y,u,\lambda) K(u,z)d\mu(u).
$$
If this can be solved analytically or numerically, then often one does not
need to compute the eigenvalues and eigenfunctions.
Alternatively, the resolvent satisfies
\begin{eqnarray}
K(y,z,\lambda)=\sum_{j=1}^\infty r_j(y)l_j(z)/(\lambda_j-\lambda).\label{res2}
\end{eqnarray}
The {\it Fredholm determinant} is
$$
D(\lambda)=\Pi_{j=1}^\infty (1-\lambda/\lambda_j)
=\exp\{-\int_0^\lambda d\lambda \int K(y,y,\lambda)d\mu(y)\}.
$$
Note that
\begin{eqnarray}
 ||{\cal K}||_2^2=\sum_{j=1}^\infty \lambda_j^{-2}.\label{k22}
\end{eqnarray}
Also since $(1-\lambda/\lambda_j)^{-1}-1=\lambda/(\lambda_j-\lambda)$,
$$ 
\lambda K(y,z,\lambda)
=\sum_{j=1}^\infty r_j(y)l_j(z)[(1-\lambda/\lambda_j)^{-1}-1].
$$
If only a finite number of eigenvalues are non-zero, the {\it kernel} $K(y,z)$
is said to be {\it degenerate}.
(For example this holds if $\mu$ puts weight only at $n$ points.)
If not, $\{l_j\}$ and  $\{r_j\}$ typically both span
$L_2(\mu)=\{ f: \int |f|^2d\mu<\infty \}.$
For $f\in L_2(\mu)$,
\begin{eqnarray}
f(y)=\sum_{j=1}^\infty R_j r_j(y)=\sum_{j=1}^\infty L_j l_j(y)
 \mbox{ where }R_j=\int fl_jd\mu, \ L_j=\int fr_jd\mu\label{L2}
\end{eqnarray}
with convergence in $L_2(\mu)$. So
\begin{eqnarray}
{\cal K}^n f(y)=\sum_{j=1}^\infty R_j r_j(y)/\lambda_j^n,\
f(y){\cal K}^n =\sum_{j=1}^\infty L_j l_j(y)/\lambda_j^n,\ n\geq 0.\label{n}
\end{eqnarray}
So if 
\begin{eqnarray}
|\lambda_1|<|\lambda_j|\label{l1}
\end{eqnarray}
 for $j>1$ then as $n\rightarrow\infty$,
$$
{\cal K}^{n+1} f(y)/{\cal K}^n f(y)\rightarrow\lambda_1^{-1},\
 f(y){\cal K}^{n+1}/ f(y){\cal K}^n \rightarrow\lambda_1^{-1}.
$$
This is one way to obtain $\lambda_1$ arbitrarily closely. Another is to use
\begin{eqnarray}
\lambda_{1}^{-1}=\sup\{ \int g {\cal K} h d\mu:\ \int ghd\mu=1\}
\mbox{ if }\lambda_{1}>0,
 \label{1}\\
\lambda_{1}^{-1}=\inf\{ \int g {\cal K} h d\mu:\ \int ghd\mu=1\}
\mbox{ if }\lambda_{1}<0.
 \label{2}
\end{eqnarray}
The maximising/minimising functions are the first eigenfunctions
$g=g_1,h=h_1$. These are unique up to a constant
multiplier if (\ref{l1}) holds.
If  $\lambda_1$ is known, one can use
$$
(\lambda_1{\cal K})^{n} f(y)\rightarrow R_1r_1(y),\
 f(y)(\lambda_1{\cal K})^{n}\rightarrow L_1l_1(y),
$$
to approximate $R_1r_1(y), L_1l_1(y)$.
Also since $l_1(y)$ is only unique up to a multiplicative constant, we may choose $R_1=1$ and so approximate  $r_1(y), l_1(y)$.
One may now repeat the procedure on the operator
${\cal K}_1$ corresponding to
$$K_1(y,z)=K(y,z)-r_1(y)l_1(z)$$
to approximate $\lambda_{2},r_{2}(y),l_{2}(z)$,
assuming the next eigenvalue in magnitude, $\lambda_{2},$ has multiplicity 1.
If say  $\lambda_{1}$  has multiplicity $M>1$, then
$$
(\lambda_1{\cal K})^{n} f(y)\rightarrow \sum_{j=1}^M R_jr_j(y),$$
and one can adapt the method above.

For further details see Withers and Nadarajah (In press).
For further details on Fredholm theory
for symmetric kernels, see Withers (1974, 1975, 1978).

\end{document}